# Nepenthes Honeypotsbased Botnet Detection

Sanjeev Kumar, Rakesh Sehgal, Paramdeep Singh
Cyber Security Technology Division, C-DAC, Mohali, INDIA
ror.sanjeev@gmail.com, rks@cdacmohali.in, paramsbhatia@gmail.com

Ankit Chaudhary
Dept. of Computer Science, BITS Pilani, INDIA
ankitc.bitspilani@gmail.com

*Abstract*— **Thenumbers of the botnet attacks areincreasing day by day and the detection of botnet spreading in the network has become very challenging. Bots are having specific characteristics incomparison of normal malware as they are controlled by the remote master server and usually don't show their behavior like normal malware until they don't receive any command from their master server. Most of time bot malware are inactive, hence it isvery difficult to detect. Further the detection or tracking of the network of theses bots requires an infrastructure that should be able to collect the data from a diverse range of data sources and correlate the data to bring the bigger picture in view.In this paper, we are sharing our experience of botnet detection in the private network as well as in public zone by deploying the nepenthes honeypots. The automated framework for malware collection using nepenthes and analysis using anti-virus scan are discussed. The experimental results of botnet detection by enabling nepenthes honeypots in network are shown. Also we saw that existing known bots in our network can be detected.**

*Index Terms*—**Malware, Bots, Network Security, Nepenthes, Honeypots, Privacy**

## I. Introduction

Botnet is becoming a major problem to internet as the size of cyber space is growing day by day. The applications running on the cyber space are becoming insecure and vulnerable to the attack generated by the black hat community. The motivation behind the attack can be to gain the access of the user's computer, to steal the information, to generate the DDoS attacks, to down the resources running in network. In last few years, the size of the network is increasing from low speed to gigabit network and the applications are also increasing day by day. In today's business oriented and heterogeneous network, security of the existing applications is extremely important. The flaws in security have become significant problems for private users, business, and even for government [1].

A botnet is a system that remotely controls malicious programs running on compromised hosts. Botnets are now a major source of network threats including DDoS, spam, identity theft, click frauds, etc. [17-19]. Botnets are still rapidly proliferating and communicating using a variety of protocols, such as IRC, HTTP, peer-to-peer, etc. The cumulative size of botnets is estimated in millions of hosts [16][18-19]. Due to the huge number of botnets, and evolving botnet protocols, it appears difficult to block or remove (or both) all bots from the Internet. So, first start with minimum target to identify bots and their actions.

The attacks by Black hat community are daily increasing on the applications running in the network to steal useful information or to gain access of the client machine. Malwares are spreading into the cyber space and bot is one of those kinds of malware which has special kind of characteristics and remotely controlled by the botmaster. By detecting the some set of attacks used by the black hat community, we would be able to tighten the security of these kinds of high speed network as well as applications running in these networks.

As described in [16], bots are typically activated by bot commands through a communication and control channel (C&C) opened by attackers (i.e. botmasters) from remote sites. The bot commandsissued may be run by a group of bots in the botnetsimultaneously, as they have been programmed. The study of bot behavior in response to issued commands is important for the development of effective countermeasures, for tracing botnet growth, and for protecting the vulnerable infrastructure which are the target of bots.Also theidentification of victims targeted by botnets may also be facilitated by a thorough analysis of bot commands [15]. More information about the bots and virus could be found out in [3-5].

Botnet detection and tracking is one of the very active research areas since last few years. Different solutions and techniques have been proposed for the same. Our approach was based on honeypot [7-8] for the malware collection and automated analysis of collected malware. A proactive system design has been discussed in [14]. The nepenthes low interaction honeypotswas used instead of high interaction honeypots. There were two main reasons of using low interaction honeypots:-

- They can be easily configured and installed. Also they are low resource intensive.
- They are much faster than high interaction honeypots.

So, with this consideration of honeypot technology, we could provide the attacker insight into our emulated network environment. Also we could monitor and log the interaction between the attackers and honeypots which could be further studied and analysed for botnet detection.

The system which was used to detect and analyse the attackers' action,was known as honeypot system. The honeypot has no intervention with the production traffic; therefore anything which comes on the honeypots is most likely the malicious intent. As compare to any other available security tools, honeypots are capable of logging much more information. They provide the vulnerable environment, to the attackers so that they can attack on the information system and can get access of the system.

Every activity of the attacker would be being monitored and logged and could be analysed. To tighten the security into the network, the nepenthes were deployed as a low interaction honeypots to study the known bot families into the network. Virtualization technology VirtualBox® [11] was used to reduce the involved hardware cost which is an open source virtualization product,which provides the flexible environment to set-up the network with single physical machine.

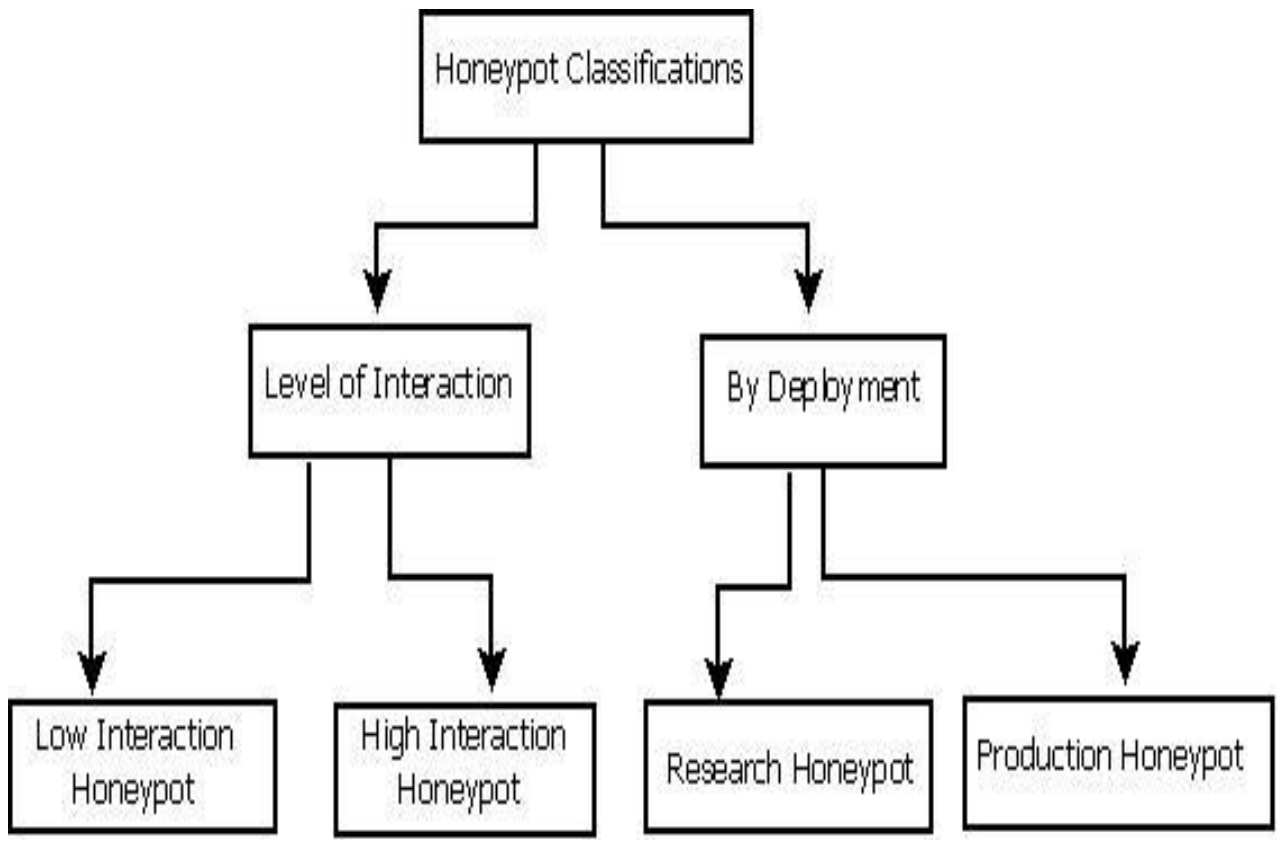

Figure1.Classification of honeypots

## II. BACKGROUND ANDTECHNOLOGY USED

A brief introduction of different technologies is following which were used in the project.

### A. Honeypot

A network security resource whose value lies in it, being scanned,attacked,compromised, controlled and misused by an attacker to attain his malicious goals. Lance Spitzner defines Honeypots as "A Honeypot is an information system resource whose value lies in unauthorized or illicit use of that resource" [1]. Honeypots can be classified into two main categories. Firstly, they can be based upon their level of interaction with an attacker. This can be further categorized as discussed in [9] [16]. Figure 1 depicts the classification of the honeypots according to their level of interaction and as per their deployment in network.

#### *1) Low-interaction honeypots*

Low interaction provides emulated network services to the attackers. Honeyd [6] and Nepenthes [13] are the examples of these kinds of low interaction honeypots. In contrast with low interaction honeypots, high interaction honeypots provides complete freedom to attackers to interact with real operating system and services and their all attemptsare logged and accounted for.

#### *2) Production Honeypots:*

They are placed withinan organization's production network for thepurpose of detection. They extend the capabilitiesof intrusion detection systems. Such honeypotsare developed and configured to integrate with theorganization's infrastructure. They are usuallyimplemented as low-interaction honeypots sittingwithin the server farm, but implementations mayvary depending on requirements of the organization.

#### *3) Research Honeypots:*

These are deployed bynetwork security researchers – the White hat hackers. They allow complete freedom for theattacker andlearntheir tactics in this process. Using research honeypots zero dayexploits, Worms, Trojans and viruses which are propagatingin the network can be isolated and studied.Researchers can then document their findings andshare them with system programmers, with networkand system administrators, with various system andanti-virus vendors. They can provide the raw materialfor the rule engines of IDS, IPS and firewallsystems.

### *B. Botnet*

A 'Bot' has special characteristics as compare to the normal malwares. They are maintained and controlled by the remote servers know as botmastes. The collection of computers infected with such bot malwares are known as botnet. Therefore botnet is a network of zombies (infected computers) which are controlled by the botmaster. Normally bot malwares are inactive and get the command from the remote server (C&C). The commands are being executed by the bot client when it is given by the C&C servers. Bot masters control the botnet through a command and controlmechanism. The formation of botnet is like C&Cservers often communicate with other C&C servers to achieve the redundancy.

The topology of a botnetevolved over time from simple star to complex random combination of different topologies.Botnets are often classified according to the protocol through which it sends out commands tothe zombie computers. A typical classification is as [2]:

- IRC Botnet: Bot masters acts as IRC servers and uses IRC channels to send commands tothe botnet. All of the members of the botnet are connected to the channel. Commands arepassed as a broadcast to the participants using the common IRC protocol.
- HTTP Botnet: Bot master acts as a web server and bots are connected to the web

server.Commands are encapsulated in HTTP messages.

- P2P Botnet: Newer breed of botnet that uses existing P2P protocols to distributecommands. This kind of botnet is harder to detect compared to the other botnets.

The bots are connected to the botnet through a C&C channel as mentioned above. A C&Cchannel can operate on different network topologies and communication mechanisms. The mostcommon protocol used for this is the IRC protocol. The main reasons why IRC is so popular are [12]:

- *Interactive*- the full two way communication between the server and the client ispossible.
- *Easy to install-setting up private servers or using existing ones are easy.*
- *Easy to control-using credentials such as username, passwords and channels; all theneeded functionalities already exist in the IRC protocol.*
- *Redundancy possibilities- by linking several servers together, one server can godown while the botnet is still functioning by connecting to other IRC servers.*

There alsoa botnet that uses the HTTP protocol for C&C. HTTP based C&C is stillcentralized, but the botmaster does not directly interact with the bots using chat likemechanisms. Instead, the bots periodically contact the C&C servers to obtain theircommands. As its proven effectiveness and efficiencies, it is expected that centralized C&C (e.g. using IRC and HTTP) will still be widely used by botnets in near future.

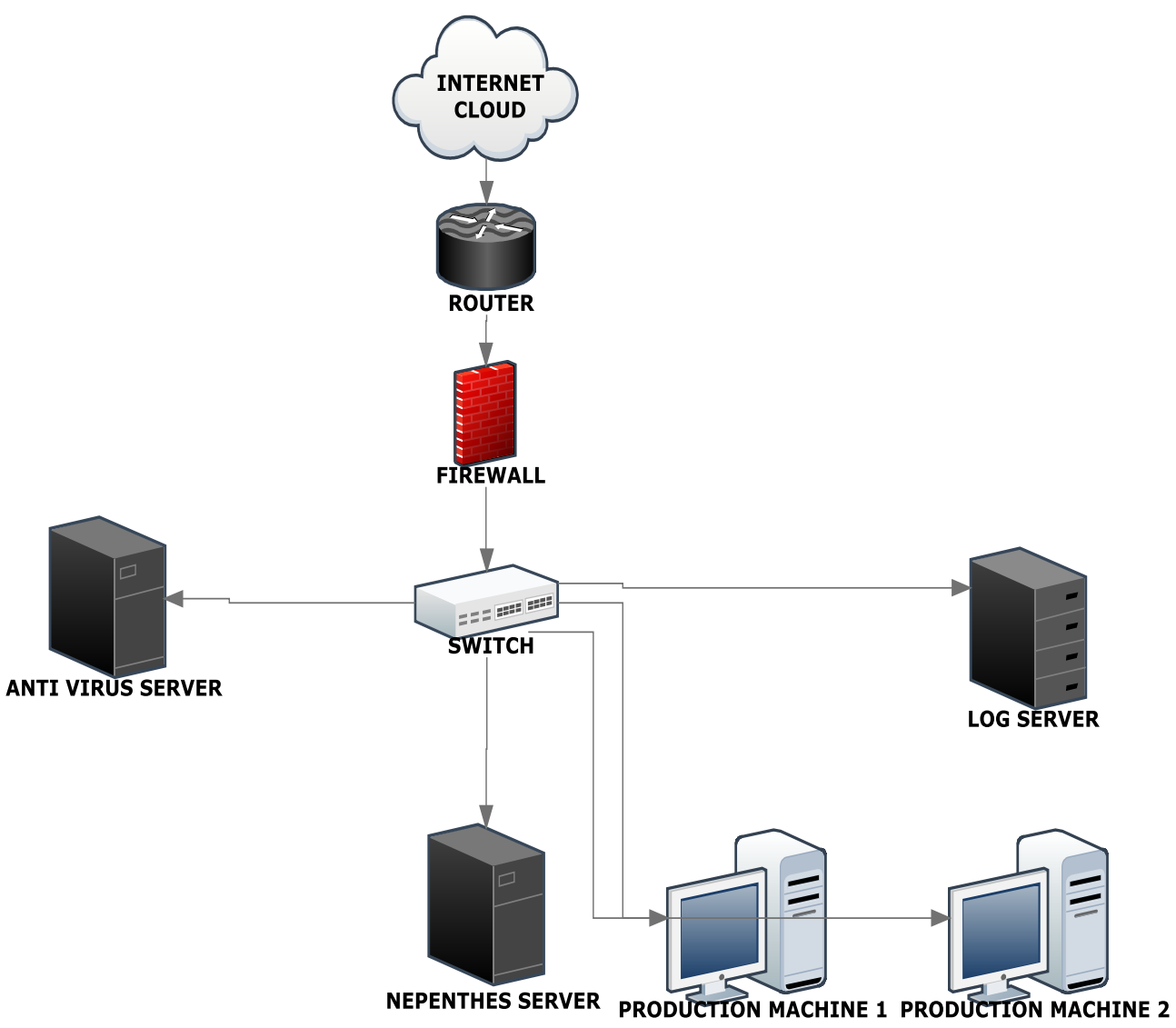

Figure 2.System Architecture

## III. SYSTEM ARCHITECTURE

The nepenthes deployment architecture is discussed for malware collection and automated analysis using anti-virus scan. Nepenthes [13][15] are low interaction honeypots widely used for malware collection. Nepenthes as low interaction honeypot with default set of vulnerabilities which can be deployed in production network and can be useful to generate the alerts to system administrator who can take the necessary actions to tighten the security of the network. We enabledthe nepenthes low interaction sensors in enterprise network and make them active all the time. We have deployed the nepenthes sensors in public network zone at various geographical locations as well as in private LAN to collect worms spreading in local private network [10].

Figure 2 show the system architecture including nepenthes and automated analysis of malwares using virus scan. As depicted in the Figure, there are nepenthes sensors used for malware collection. Basically there are three major components of the system: Malware collector, Virus scan server and Log server.

### *a) Malware Collector*

As shown in the Figure 2, the low interaction honeypot (nepenthes) was used for malware collection. This is discussed how a particular low-interaction honeypot (nepenthes) [13] could be used to quickly alert an administrator about a network compromise. It captures malware and can assist in containing and removing the infection.

### *b) Nepenthes Sensors*

For the implementation of nepenthes sensors VirtualBox® [11] was used. By using virtualization, it will reduce the hardware cost as compare to real physical system as well as will improve the deployment and maintenance.

Various Modules in Nepenthes

- *Vulnerability Modules* – emulates various services which look ripe for compromise to an attacker (lsass, dcom, veritas, dameware, etc)
- *Shellcode Handlers and Emulators* – allows nepenthes to interact with the malware
- *Download Modules* – will download the binary (http, ftp, curl, etc)
- *Submission Modules* – will submit the binary for analysis (Norman, CWSandbox, postgres, etc)

### *c) Log server*

Malware collected and all the data sets including network traces, pcap data were stored in log server for further analysis of the collected data. Log server is a central database server which keeps the metadata of the collected information. It keeps the following records:

- MD5 values of the malware samples
- Malware Binaries
- Pcap data & network traces
- Analysis results including antivirus labels etc
- IP address information
- Logs of the download, submitted binaries

Malware binaries stored in log servers were fetched and submitted to the analysis server where further analysis of the corresponding binary was done and result logs were putted into log server.

*d) Anti-Virus Scan:*

The malware binaries were fetched from log server and automatically submitted to anti-virus scan server which was doing the analysis of the binary based on predefined signatures. For this purpose three antivirus softwarewere chosen from different companiesMacAfee®, Symantec® and Microsoft®. Also the MD5 of the corresponding binarywas submitted to theVirus Total [15] for scan with 42 anti-virus products. Virus Total is a free online service that enables Internet users to scan dubious files with 42 different antivirus (AV) tools.

The functionality of the system is as the following: the user sends a file to the system, via email or the web interface. He would get a report back when all AV tools will have finished examining the submitted file. That report includes the output of each engine, URLs with extra information about the potential threat if any. It gives information about file metadata size, various hashes of the file etc. It can also contain packet identification or the Portable Executable (PE) structure information of the malware. Virus Total with its 42 AV engines, offers a valuable service not only to the end users but also to the community of the AV vendors. Indeed, Virus Total can provide them with samples of malware that match certain criteria of interest to them. In the general case Virus Total sends a malware sample to AV vendor X then the following would be done:

- If at least one other AV engine has detected the sample as being malicious whereas the AV engine of X has not.
- If the AV engine from X has detected that sample as being malicious using a generic pattern or a heuristic.

Most AV vendors follow these two rules but some of them impose other criteria also. For instance, some have decided to get samples that are detected by at least N out of K AV engines and that their own has missed. Others do restrict even further the conditions by imposing that all engines from a well-defined subset of engines must have detected the sample and that their own has missed it [13]. Clearly the amount of samples to be sent to the AV vendors is a function of the filtering rules they have chosen. It is worth noting though that, in the general case, some vendors do get as many as 10000 samples per day [15].However this kind of malware collection mechanism may incur more cost and require complete collaboration with antivirus vendors in terms of services. This solution would be good for large organization, individual researchers, small organization, and private partners.

The complete process of malware collection and their analysis can be represented in following way:

1. If the nepenthes honeypots are deployed in public network zone, then there is central malware collection repository of the malware.
2. If the nepenthes honeypots are deployed in private network then there is local malware repository and analysis server

The system is working on 3 layer architecture. Layer 1 incorporates nepenthes honeypot sensors which captures the malware samples and sends data to the central server on a regular basis. Layer2 incorporates central server which performs activities like registering new nepenthes nodes, processing data sent by remote nodes by fusing the data with the configuration information of honeypots and converting the data in to a relational data base format. Layer 3 consist the database which acts as a data source for analysis engine. Figure 3 depicts the complete flow and deployment of the system.

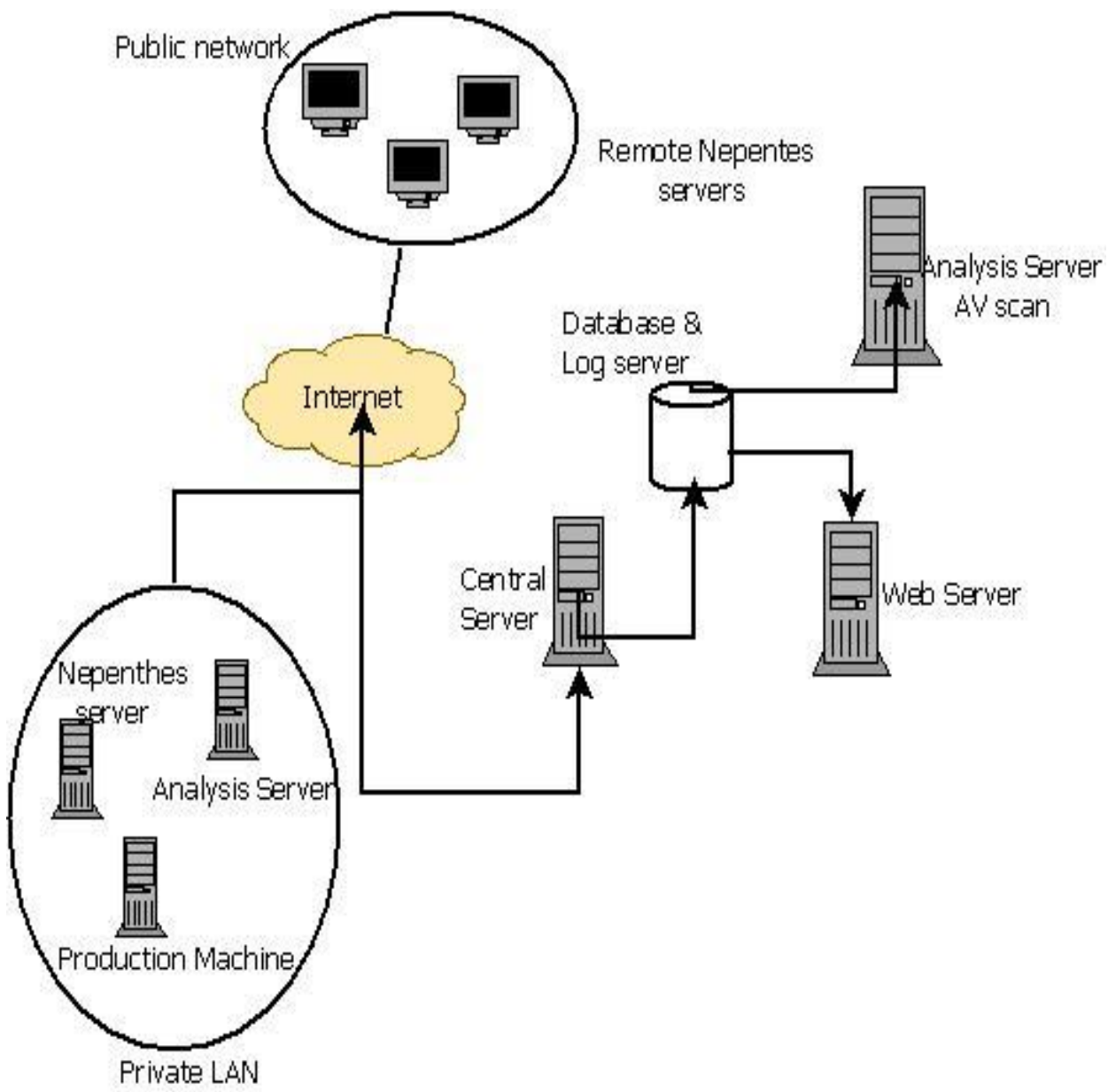

Figure 3.Complete Flow of the System

## IV EXPERIMENTAL RESULTS

Here we are presenting some results of our nepenthes based system as malware collector and analysis mechanism. We have implemented our mentioned system on 1/1/2011 and collected very valuable information and malwares. We have collected real bots which were damaging the computers in network. During this period of deployment total 732 numbers of samples were collected and we are containing large amount of PCAP data which is highly malicious in nature. Continuously we are submitting the data to our centre response team which are taking the remedial actions against the collected data sets and corresponding IP or attackers. Below Figure 4 illustrates the top 10 countries from where we have collected most our data sets.

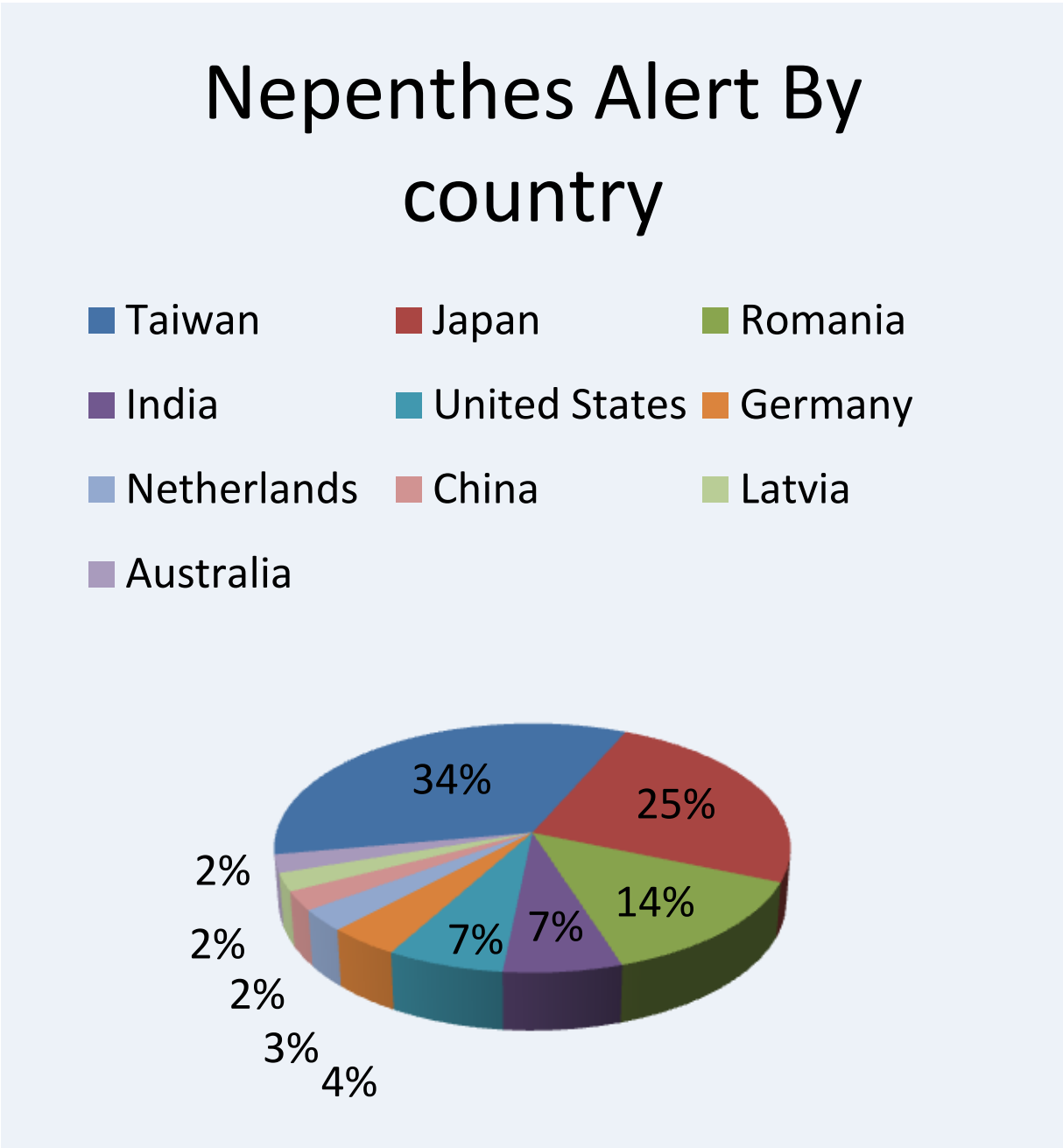

Figures 4.Top 10 alerts from different countries

Below TableIshow the top 10 most accessed URLs by deploying the nepenthes sensors geographically. Column 1 in Tablerepresent the URLs accessed and column 2 represent the hit count of the corresponding URL. As shown in Table the hit count of the www.x.x.x.de/M.txt is 191 which are highest among others. For security concern, we have changed the name of the URLs.

TABLE I.
MOST ACCESSED URLS

| URL | Count |
|---|---|
| http://www.x.x.x.de/M.txt | 191 |
| tftp://x.x.x.x/host.exe | 97 |
| http://y.net/fastenv | 39 |
| http://nmap.org/book/nse.html | 24 |
| http://y.proxyfire.net/fastenv | 19 |
| http://XX.63.156.12:8326/lsd | 18 |
| http://www.a.a.com | 17 |
| http://XXX.109.153.3/proxycheck.txt | 15 |
| http://XXX.109.153.5:11111 | 15 |
| tftp://XXX.100.78.32/host.exe | 14 |

TableII shows the top 10 MD5 values and their labelling corresponding to anti-virus scan. Column 1 represents the MD5 value of the collected malware binary, column 2 depicts hit count, and column 3, 4 and 5 depicts the antivirus labelling corresponding to MacAfee®, Microsoft®and Symantec® antivirus products. As we can see most of them are declared as IRC W32.IRC bot malwares. When we have submitted MD5 of that binary to Virus Total for scanning with 42 antivirus products, they were really the IRC bots and most of the antivirus products declaring them as IRC bots. So our deployed our automated system easily detected the bots spreading in the network and tighten the security against these bots.

TABLE II.
TOP 10 MD5 VALUES & THEIR LABELLING

| MD5 | Count | MacAfee | Microsoft | Symantec |
|---|---|---|---|---|
| 8659156 50a85e7 c27cdd1 1850a13 f86e | 51 | W32/Sdb ot.worm. gen.bs | BackdoorWin32 /Rbot | W32.IRCB ot |
| 809fe9b 32845ed f5c09b8 71e0e68 f227 | 63 | W32/Sdb ot.worm. gen.bs | BackdoorWin32 /Rbot | W32.IRCB ot |
| 0da155b 04f16daf afffbb1a 485b3d0 e1 | 27 | W32/Sdb ot.worm. gen.bs | BackdoorWin32 /Rbot | W32.IRCB ot |
| 6e2fa90 31a05b9 649da06 2c550d1 4a3d | 40 | W32/Sdb ot.worm. gen.bs | BackdoorWin32 /Rbot | W32.IRCB ot |
| f9dc394 5bdd740 6bd8db0 6a47963 ec14 | 67 | W32/Sdb ot.worm. gen.bs | BackdoorWin32 /Agent | |
| 8a5ce07 df6a535 7dafa84f 5317aaa d35 | 75 | W32/Sdb ot.worm. gen.bs | BackdoorWin32 /Rbot | W32.IRCB ot |
| 9019b23 f2a5a51c 3367173 9af2f309 92 | 32 | W32/Sdb ot.worm. gen.bs | BackdoorWin32 /Rbot | W32.IRCB ot |
| 15965bb 88165d1 eb06851 d8f0761 30ba | 31 | W32/Sdb ot.worm. gen.bs | BackdoorWin32 /Rbot | W32.IRCB ot |

Also Figure 5 shows the daily collection graph by our system since its deploying date.

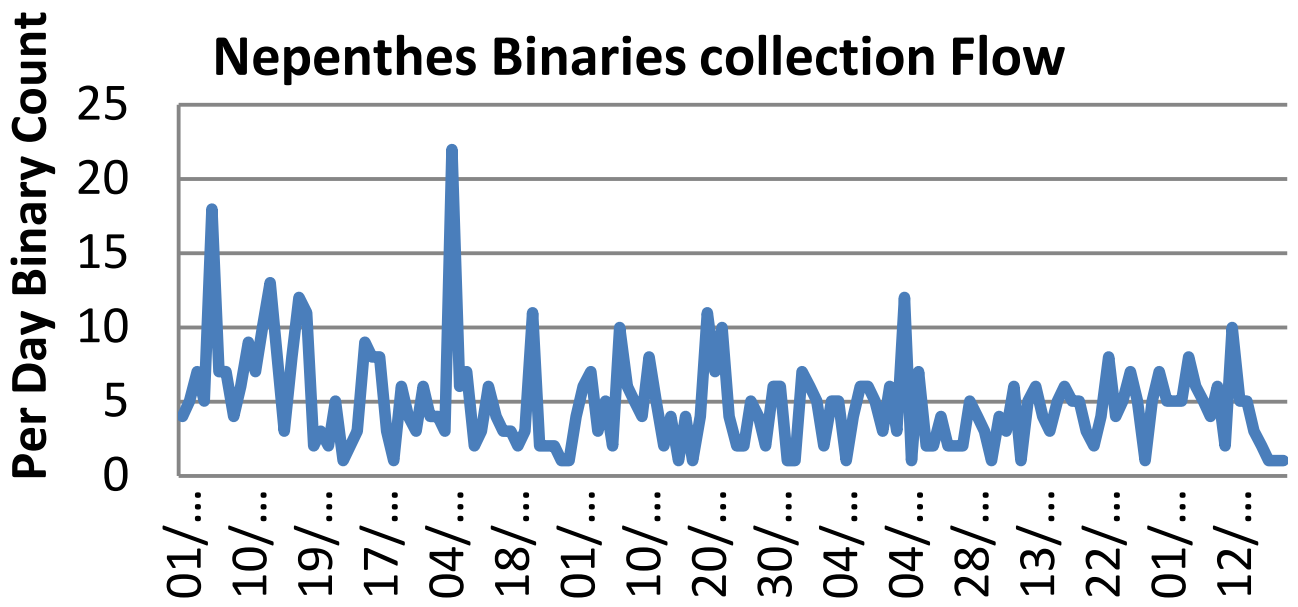

Figure5. Nepenthes Binaries Daily Collection

Some results of logs generated on the nepenthes honeypots are shown below when deployed with public IPs. Nepenthes honeypot IP address was 203.x.x.x and others were the foreign outside IP addresses. As we can see there is binary download ftp andtftp protocol. These results signify the interaction of the outside IP address with nepenthes as honeypots don't have any production values which clarify that they are malicious IP addresses. For security reason we have omitted the real IP address.

```
[2011-05-27T15:53:16]    66.x.x.x    ->
203.x.x.x creceive://66.x.x.x:9988/0
[2011-05-27T16:56:59]    59.x.x.x    ->
203.129.220.217
creceive://59.x.x.x:9988/0
[2011-05-28T04:57:15]    83.x.x.x    ->
203.x.x.x creceive://83.x.x.x:9988/0
[2011-05-28T12:03:57]    203.x.x.x   ->
203.x.x.x
tftp://203.111.222.65/host.exe
[2011-05-28T16:26:30]    203.x.x.x   ->
203.x.x.x tftp://203.x.x.x/host.exe
[2011-05-28T21:21:11]    60.x.x.x    ->
203.x.x.x creceive://60.x.x.x:9988/0
[2011-05-29T02:52:18]    203.x.x.x   ->
203.x.x.x
ftp://1:1@203.x.x.x:12405/host.exe
[2011-05-29T06:34:37]    203.x.x.x   ->
203.x.x.x tftp://203.x.x.x/host.exe
[2011-05-29T08:53:48]    209.x.x.x   ->
203.x.x.x creceive://209.x.x.x:9988/0
[2011-05-30T01:08:45]    174.x.x.x   ->
203.x.x.x creceive://174.x.x.x:9988/0
[2011-05-30T02:54:41]    125.x.x.x   ->
203.x.x.x http://www.baidu.com/
[2011-05-30T06:28:48]    203.x.x.x   ->
203.x.x.x tftp://203.x.x.x/host.exe
[2011-05-30T06:47:21]    203.x.x.x   ->
203.129.220.217
ftp://1:1@203.180.24.32:18807/host.exe
```

## V. Conclusion and Future Work

In this paper we have presented anautomated system based on nepenthes as malware collector and analysis of them using antivirus scan. This is one step of detecting the known bots in the network and in any organization we can detect the bot spreading in the network using this system. Further we have also shown the behavior based analysis of the collected malwares which are not detected by the antivirus products. The claim is that all the software (OS and associated tools) are Open Source. A low-interaction honeypot like nepenthes is easy to install and requires minimal maintenance.It may provide valuable information in the event of an infection within your organization. Whenused in conjunction with an Intrusion Detection System, valuable information about thebehavior of the malware, packet captures and the malware binary itself may be obtained.

## Acknowledgment

We would like to thank Cyber Security Technology team at C-DAC, Mohali to provide the infrastructure and recourses to collect the malwares and to available them for further analysis. We are also very thankful to Executive Director of CDAC, Mohali to provide us full support. This research was supported by DST, Ministry of Science & Technology, Govt. of INDIA.

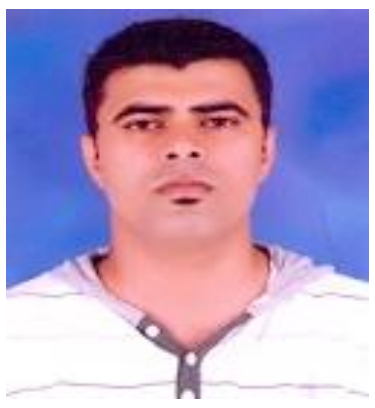

Sanjeev Kumar received his B.Tech from Kurukshetra University, INDIA and pursuing M.Tech in CS from PTU INDIA. He is working as a staff scientist at CDAC, Mohali, INDIA. He is CCNA and CCNP certified and an active member of Indian National Grid known as GARUDA. His technical expertises are in networking, network security.

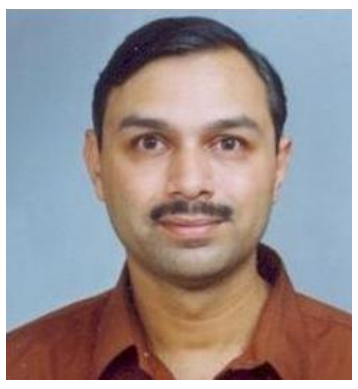

RakeshSehgal received his B.E in Electronics from Nagpur University, 1988 and M. Tech. in Computer Science from DAU, Indore. He is currently Principal Design Engineer & Head of Cyber Security Technology Division at CDAC- Mohali. He has vast research experience in Network Security, Honeynets and Honeypots.

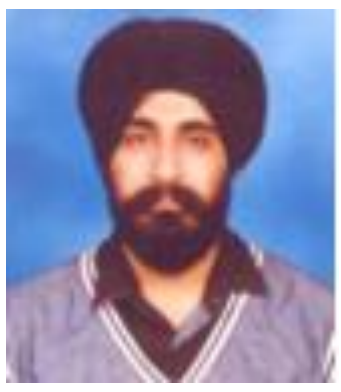

Paramdeep Singh received his MCA from PTU Punjab INDIA. Currently he is working in Cyber Security Technologies Division at CDAC Mohali, INDIA. He has extensive work experience on Honeynets and Honeypots.

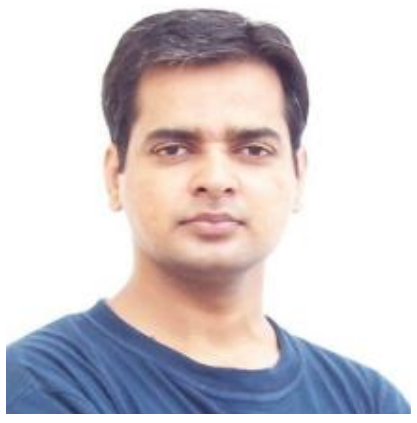

Ankit Chaudhary received his ME & PhD in CSE from Birla Institute of Technology & Science, Pilani, INDIA. His research interests are Machine Learning, Artificial Intelligence, Network Security and Mathematical Computations.